\documentstyle[aps,multicol,prb,epsf]{revtex} 
\begin{document} 
%\draft 

\title{Magnon-assisted transport and thermopower in
ferromagnet-normal metal tunnel junctions} 
\author{Edward McCann and Vladimir I. Fal'ko} 
\address{Department of Physics, Lancaster University,
Lancaster, LA1 4YB, United Kingdom}
\date{\today} 
\maketitle
\begin{abstract}
{We develop a theoretical model of magnon-assisted transport in a mesoscopic
tunnel junction between a ferromagnetic metal and a normal (non-magnetic) metal.
The current response to a bias voltage is dominated by the contribution
of elastic processes rather than magnon-assisted processes
and the degree of spin polarization of the current, parameterized by a
function $P\;\!\!( \Pi_{\uparrow (\downarrow )},\Pi_{N} )$, $0 \leq P \leq 1$,
depends on the relative sizes of the majority $\Pi_{\uparrow}$ and
minority $\Pi_{\downarrow}$ band Fermi surface in the ferromagnet and of the
Fermi surface of the normal metal $\Pi_{N}$.
On the other hand, magnon-assisted tunneling gives the dominant contribution
to the current response to a temperature difference across the junction.
The resulting thermopower is large,
$S \sim - (k_B/e) (k_BT/\omega _{D})^{3/2}
P \;\!\! ( \Pi_{\uparrow (\downarrow )},\Pi_{N} )$,
where the temperature dependent factor $(k_{B}T/\omega _{D})^{3/2}$
reflects the fractional change in the net magnetization of the ferromagnet
due to thermal magnons at temperature $T$ (Bloch's $T^{3/2}$ law) and
$\omega _{D}$ is the magnon Debye energy.
}\end{abstract}

 
\begin{multicols}{2}
\bibliographystyle{simpl1}

%%%%%%%%%%%%%%%%%%%%%%%%%%%%%%%%%%%%%%%%%%%%%%%%%%%%%%%%%%%%%%%%%%%%
%%%%%%%%%%%%%%%%%%%%%%%%%%%%%%%%%%%%%%%%%%%%%%%%%%%%%%%%%%%%%%%%%%%%
\section{Introduction}
%%%%%%%%%%%%%%%%%%%%%%%%%%%%%%%%%%%%%%%%%%%%%%%%%%%%%%%%%%%%%%%%%%%%
%%%%%%%%%%%%%%%%%%%%%%%%%%%%%%%%%%%%%%%%%%%%%%%%%%%%%%%%%%%%%%%%%%%%

For more than a decade there has been intensive research of spin polarized
transport\cite{pri95} with a view to exploit phenomena such as the tunneling
magnetoresistance (TMR) of ferromagnetic junctions\cite{jul75,moo00} and the
giant magnetoresistance (GMR) in multilayer structures.\cite{bai88,pra91}
The TMR effect arises because the mismatch of spin currents at the interface
between two ferromagnetic (F) electrodes with antiparallel spin polarizations
produces a larger contact resistance than a junction with parallel
polarizations.
In general, the magnitude of TMR increases with the degree of polarization
of the ferromagnets, but it is reduced by spin relaxation
phenomena\cite{pri95,j+s85},
such as spin-orbit scattering at impurities or magnon emission,
which lift spin current mismatch.
Magnon emission is an inelastic process that has been studied both
theoretically\cite{bra98} and experimentally\cite{tsu71} in ferromagnetic
tunnel junctions with a view to relate nonlinear $I(V)$ characteristics
to the density of states of magnons $\Omega (\omega)$
as $d^2I/dV^2 \propto \Omega (eV)$.
As well as influencing the magnetoresistance, magnon emission is expected to
produce a magnetothermopower effect: a larger thermopower in the antiparallel
orientation than in the parallel one.\cite{m+f02a,m+f02b}

Meanwhile, interest in transport across junctions between ferromagnets and
normal metals or semiconductors has focused on the possibility of injecting
spin into the normal system.\cite{j+s85,aro76,d+d90}
Current in the ferromagnet (F) is carried unequally by majority 
and minority carriers so that a current flowing across the interface with
a normal conductor is expected to have a finite degree of spin
polarization $P$, $0 \leq P \leq 1$.
Spin injection from a ferromagnetic metal into a normal metal (N)
has been measured in broad agreement with theory,\cite{j+s85,aro76}
but there has been difficulty in achieving large degrees of spin injection
into a semiconductor at room temperature.\cite{lee99}
A limiting factor is believed to be conductivity mismatch,\cite{sch00}
a problem that may be overcome by introducing a tunneling barrier between
the ferromagnet and semiconductor.\cite{ras00}
Spin injection in all-semiconducting devices has generally been more
successful,\cite{fie99}
although it is limited to relatively low temperatures at the moment.

In this paper we investigate the opposite effect: the injection of charge
caused by the equilibration of magnetization between ferromagnetic baths
kept at different temperatures. 
The magnetization of a ferromagnet held at a finite temperature $T$
is less that its maximum value due to the thermal occupation of magnons,
with the fractional change $\delta m(T)$ obeying Bloch's $T^{3/2}$ law,
\begin{eqnarray}
\delta m(T) &=& \frac{3.47}{\xi} 
\left( \frac{k_BT}{\omega_D} \right)^{3/2} , \label{mt0} 
\end{eqnarray}
where $\xi$ is the spin of the localized moments
and $\omega_D$ is the magnon Debye energy.
Under certain conditions the transfer of heat across a junction between
ferromagnets or a ferromagnet and a normal metal may involve
simultaneous spin transfer and, possibly, charge transfer.
Therefore we study the influence of magnon assisted processes on
the transport properties of a mesoscopic size
ferromagnet/insulator/normal metal tunnel junction.
The bottle-neck of both charge and heat transport lies in a small-area tunnel
contact between the electrodes held at different temperatures
and/or electric potentials.
The main result is a large thermopower $S$ arising from the
magnon-assisted processes that depends on the difference between the size
of the majority and minority band Fermi surfaces in the ferromagnet,
\begin{eqnarray}
S \approx - \frac{k_B}{e}
\frac{\delta m(T)}{2} 
P \;\!\!\left( \Pi_{\uparrow (\downarrow )},\Pi_{N} \right) . \label{Sintro} 
\end{eqnarray}
This result holds in a range of temperatures given by
\begin{eqnarray}
1 \gg \delta m(T)
\gg (k_B T)/\epsilon_F ,
\label{reg0}
\end{eqnarray}
where $\epsilon_F$ is the Fermi energy. The function $P$, $0 \leq P \leq 1$,
is the degree of spin polarization of the current response to a bias voltage
and it depends, in general, on the area of the maximal cross-section
of the Fermi surface in the plane parallel to the interface
of majority $\Pi_{\uparrow}$ and minority $\Pi_{\downarrow}$ electrons
in the ferromagnet, and of the electrons in the normal (non-magnetic)
metal $\Pi_{N}$.
As an example,
\begin{eqnarray}
P &=& \frac{\left( \Pi_{N} - \Pi_{\downarrow} \right)}
{\left( \Pi_{N} + \Pi_{\downarrow} \right)} ,
\label{pol}
\end{eqnarray}
for a momentum-conserving tunneling model
with $\Pi_{\uparrow} \geq \Pi_{N} > \Pi_{\downarrow}$.

The magnon-assisted processes that we consider are similar to those
discussed in Refs.~\onlinecite{bra98,m+f02a,m+f02b} in
relation to transport in F-F tunnel junctions.
Microscopically, a typical magnon-assisted process that contributes to
the thermopower in a F-N junction, Eq.~(\ref{Sintro}), is shown
schematically in Figure~1.
Here the majority electrons in the ferromagnet on the left hand side of the
junction are `spin-up' while the minority electrons are `spin-down'.
In the normal (non-magnetic) metal on the right, the density of states of
the spin-up and spin-down conduction electrons are equal.
The transition begins with a spin-down electron on the right,
that then tunnels through the barrier (without spin flip) into an
intermediate, virtual state with spin-down minority polarization on the left
(Figure~1(a)).
In the final step, Figure~1(b), the electron emits a magnon
and incorporates itself into a previously unoccupied state in the
spin-up majority band on the left.

In our approach, we take into account inelastic tunneling processes
that involve magnon emission and absorption in the ferromagnet,
as well as elastic electron transfer processes,
in order to obtain a balance equation for the current $I(V,\Delta T)$ as a
function of bias voltage, $V$, and of the temperature drop, $\Delta T$.
In the linear response regime the electrical current may be written as
\begin{eqnarray}
I = G V + L \, \Delta T ,
\label{gvgt}
\end{eqnarray}
where $G$ is the electrical conductance and $L$
describes the response to a temperature difference.
Under conditions of zero net current, the thermopower coefficient is
\begin{eqnarray}
S = - \frac{V}{\Delta T} = \frac{L}{G} .
\label{defS}
\end{eqnarray}
The electrical conductance is dominated by a contribution
from elastic processes $G \approx G_{\rm el}$ that involve tunneling
from both the majority and minority bands of the ferromagnet to the
conduction band of the normal metal
(and vice versa) without spin flip scattering.
However the contribution of the same processes to the $\Delta T$ response
is generally small,
$L_{\rm el} \sim (k_B/e) (T/\epsilon_F) G_{\rm el}$,
as it contains the additional parameter $T/\epsilon_F$.
By way of contrast, we find that responses arising from magnon-assisted
transport are of the same order,
$L_{\rm in} \sim (k_B/e) G_{\rm in} \sim (k_B/e) \delta m(T) G_{\rm el}$,
both containing the parameter $\delta m(T)$.
The result, in the temperature regime defined above in Eq.~(\ref{reg0}),
is that the overall $\Delta T$ response is dominated by inelastic processes
$L \approx L_{\rm in} \gg L_{\rm el}$ producing a large thermopower
$S \approx L_{\rm in}/G_{\rm el}$, Eq.~(\ref{Sintro}).

The paper is organised as follows.
In Section~\ref{SECTmodel} we introduce the model and technique used for
describing transport across a tunnel junction and in Section~\ref{SECTcurrent}
we calculate the current including the contribution of elastic processes
and the influence of magnon-assisted processes.
Section~\ref{SECTthermo} gives the resulting thermopower for two different
models of the interface: a uniformly transparent interface where the component
of momentum parallel to the interface is conserved, and a randomly transparent
interface.

%
%%%%%%%%%%%%%%%%%%%%%%%%%%%%%%%%%%%%%%%%%%%%%%%%%%%%%%%%%%%%%%%%%%%%%%%%%%
%%%%%%%%%%%%%%%%%%%%%%%%%%%%%%%%%%%%%%%%%%%%%%%%%%%%%%%%%%%%%%%%%%%%%%%%%%
%%%%%%%%%%%%%%%%%%%%%   figure 1 is inserted HERE %%%%%%%%%%%%%%%%%%%%%%%%
%%%%%%%%%%%%%%%%%%%%%%%%%%%%%%%%%%%%%%%%%%%%%%%%%%%%%%%%%%%%%%%%%%%%%%%%%%
%%%%%%%%%%%%%%%%%%%%%%%%%%%%%%%%%%%%%%%%%%%%%%%%%%%%%%%%%%%%%%%%%%%%%%%%%%
\begin{figure}
\hspace{0.05\hsize}
\epsfxsize=0.8\hsize
\epsffile{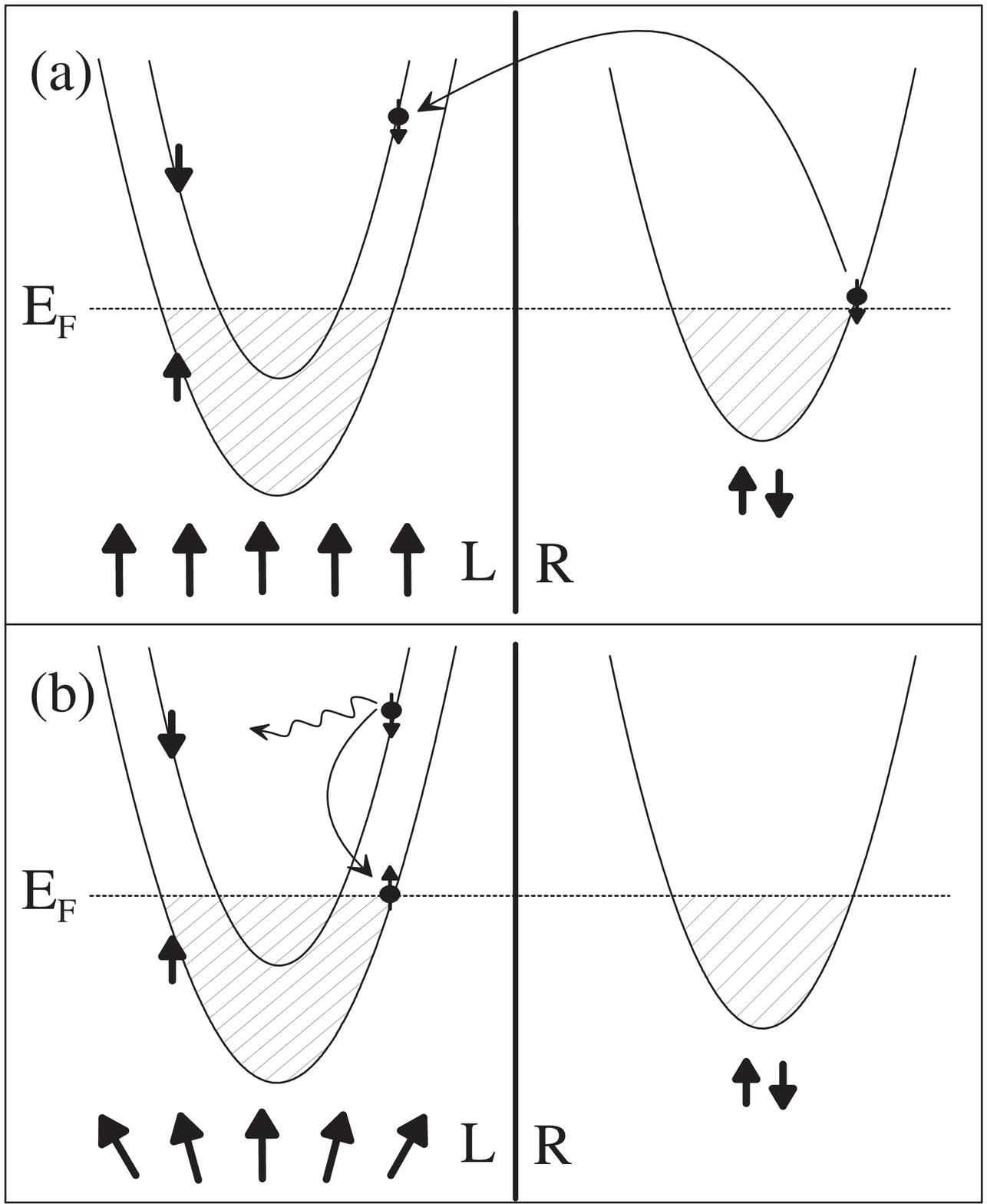}
\refstepcounter{figure}
\label{figure:1}

{\setlength{\baselineskip}{10pt} FIG.\ 1.
Schematic of magnon-assisted tunneling via an intermediate minority state
between a ferromagnet on the left and a normal metal on the right.
This example shows a transition from an initial spin-down state on the right
in the normal metal to a final majority spin-up state on the left
in the ferromagnet.
(a) The process begins with a
spin-down electron on the right, that then tunnels
through the barrier (without spin flip) into an intermediate, virtual state
with spin-down minority polarization on the left.
(b) The electron emits a magnon (wavy line) and incorporates itself into
a previously unoccupied state in the spin-up majority band on the left.
}
\end{figure}
%%%%%%%%%%%%%%%%%%%%%%%%%%%%%%%%%%%%%%%%%%%%%%%%%%%%%%%%%%%%%%%%%%%%%%%%%%
%%%%%%%%%%%%%%%%%%%%%%%%%%%%%%%%%%%%%%%%%%%%%%%%%%%%%%%%%%%%%%%%%%%%%%%%%%
%%%%%%%%%%%%%%%%%%%%%   figure 1 is inserted above %%%%%%%%%%%%%%%%%%%%%%%
%%%%%%%%%%%%%%%%%%%%%%%%%%%%%%%%%%%%%%%%%%%%%%%%%%%%%%%%%%%%%%%%%%%%%%%%%%
%%%%%%%%%%%%%%%%%%%%%%%%%%%%%%%%%%%%%%%%%%%%%%%%%%%%%%%%%%%%%%%%%%%%%%%%%%
%

%
%%%%%%%%%%%%%%%%%%%%%%%%%%%%%%%%%%%%%%%%%%%%%%%%%%%%%%%%%%%%%%%%%%%%
%%%%%%%%%%%%%%%%%%%%%%%%%%%%%%%%%%%%%%%%%%%%%%%%%%%%%%%%%%%%%%%%%%%%
\section{Description of the model}\label{SECTmodel}
%%%%%%%%%%%%%%%%%%%%%%%%%%%%%%%%%%%%%%%%%%%%%%%%%%%%%%%%%%%%%%%%%%%%
%%%%%%%%%%%%%%%%%%%%%%%%%%%%%%%%%%%%%%%%%%%%%%%%%%%%%%%%%%%%%%%%%%%%
%

We consider a tunnel junction between a ferromagnetic metal on the left ($L$)
and a normal (non-magnetic) metal on the right ($R$) with, in general,
a temperature drop $\Delta T$ and a bias voltage $V$ across the junction.
Our initial aim is to write a balance equation for the current $I(V,\Delta T)$
in terms of the occupation numbers of electrons
$n_{L(R)} (\epsilon_{{\bf k}\alpha}^{L(R)}) =
[\exp ((\epsilon_{{\bf k}\alpha}^{L(R)}
 - \epsilon_F^{L(R)})/(k_BT_{L(R)})) + 1]^{-1}$
on the left (right) hand side of the junction
and of magnons
$N_{L} ({\bf q}) =
[\exp (\omega_{\bf q}/(k_BT_{L})) - 1]^{-1}$
in the ferromagnet.
Here $T_{L(R)}$ is the temperature on the left (right) hand side,
$\epsilon_F^{L} - \epsilon_F^{R} = -eV$,
and $\omega_{\bf q}$ is the energy of a magnon of wavevector ${\bf q}$.
In the following we set $T_L = T + \Delta T$ and $T_R = T$
and we shall speak throughout in terms
of the transfer of electrons with charge $-e$.
The index $\alpha = \{ \uparrow , \downarrow \}$ takes account of the
splitting of conduction band electrons into
`spin-up' $\epsilon_{{\bf k}\uparrow}^{L(R)}$ and
`spin-down' $\epsilon_{{\bf k}\downarrow}^{L(R)}$ subbands.
We assume that the majority electrons in the ferromagnet are spin-up
so that $\epsilon_{{\bf k}\uparrow}^L = \epsilon_{\bf k}^L - \Delta/2$ and 
$\epsilon_{{\bf k}\downarrow}^L = \epsilon_{\bf k}^L + \Delta/2$
where $\epsilon_{\bf k}^L$ is the bare electron energy and
$\Delta$ is the spin splitting energy.
In the non-magnetic metal on the right, 
$\epsilon_{{\bf k}\uparrow}^R =
\epsilon_{{\bf k}\downarrow}^R = \epsilon_{\bf k}^R$.

The total Hamiltonian of the system is
\begin{eqnarray}
H &=& H_F^L + H_N^R + H_T ,  \label{Htot} \\
H_N^{R} &=& \sum_{{\bf k}\alpha} \epsilon_{{\bf k}}^R \,
c^\dag_{{\bf k}\alpha}c_{{\bf k}\alpha}, \label{hR}  \\
H_{T} &=&
\sum_{{\bf k k^{\prime}}\alpha}
\left[ t_{{\bf k},{\bf k^{\prime}}}
c_{{\bf k}\alpha}^{\dag} c_{{\bf k^{\prime}}\overline{\alpha}}
+ t_{{\bf k},{\bf k^{\prime}}}^{*}
c_{{\bf k^{\prime}}\overline{\alpha}}^{\dag} c_{{\bf k}\alpha} \right] ,
\label{ht} 
\end{eqnarray}
where $H_N^R$ is the Hamiltonian describing the conduction band
electrons of the non-magnetic metal on the right written in terms
of creation and annihilation Fermi operators $c^{\dag}$ and $c$.
The term $H_T$ is the tunneling Hamiltonian,\cite{coh62,car72,mah81}
$\overline{\alpha} \equiv \{ \downarrow , \uparrow \}$ and
we assume that spin is conserved when an electron tunnels across the interface.
The tunneling matrix elements $t_{{\bf k},{\bf k^{\prime}}}$ describe the
transfer of an electron with wavevector ${\bf k}$ on the left
to the state with ${\bf k^{\prime}}$ on the right.
In our model\cite{mcc01,m+f02b} we neglect its explicit energy dependence,
but describe both clean and diffusive interfaces by taking into account
whether the component of momentum parallel
to the interface is conserved.

The term $H_F^{L}$ is the Hamiltonian of the ferromagnetic electrode
on the left side of the junction in the absence of tunneling. 
We use the so called s-f (s-d) model,\cite{w+w70,nag74} which assumes that
magnetism and electrical conduction are caused by different
groups of electrons that are coupled via an intra-atomic exchange interaction,
although we note that the same results, in the lowest order of
electron-magnon interactions, may be obtained from a model of itinerant
ferromagnets.\cite{e+h73}
The magnetism originates from inner atomic 
shells (e.g., d or f) which have unoccupied electronic orbitals and,
therefore, possess magnetic moments whereas the 
conduction is related to electrons with delocalized wave functions.
Using the Holstein-Primakoff transformation\cite{h+p40} the 
operators of the localized moments in the interaction Hamiltonian
can be expressed via magnon creation and 
annihilation  operators $b^\dag,  b$.
At low temperatures, where the average number of magnons is small
$<b^\dag  b>\ll 2\xi$ ($\xi$ is the spin of the localized moments),
the Hamiltonian of the ferromagnet $H_F^{L}$ can be written as follows
\begin{eqnarray}
H_F^{L} = H_e^{L} + H_m^{L} + H_{em}^{L},
\label{hf} 
\end{eqnarray}
\begin{eqnarray}
H_e^{L} = \sum_{{\bf k}\alpha}\epsilon_{{\bf k}\alpha}^{L}
c^\dag_{{\bf k}\alpha}c_{{\bf k}\alpha},
\qquad 
\epsilon_{{\bf k}\uparrow ( \downarrow )}^{L} =
\epsilon_{\bf k}^{L} \mp \Delta/2,
\label{he}  
\end{eqnarray}
\begin{eqnarray}
H_m^{L} = \sum_{{\bf q}} \omega_{{\bf q}} b^{\dag}_{{\bf q}} b_{{\bf q}},
\qquad \omega_{q=0}=\omega_0,
\label{hm}  
\end{eqnarray}
\begin{eqnarray}
H_{em}^{L} = \!
- \frac{\Delta}{\sqrt{2 \xi\cal{N}}} \!
\sum_{{\bf k q}} \!
\left[ c_{{\bf k-q}\uparrow}^{\dag} c_{{\bf k}\downarrow} b_{{\bf q}}^{\dag}
+ c_{{\bf k}\downarrow}^{\dag} c_{{\bf k-q}\uparrow} b_{{\bf q}} \right] , \!\!
\label{hem}
\end{eqnarray}
The first term $H_e^{L}$, Eq.~(\ref{he}),
deals with conduction band electrons which are split into majority 
$\epsilon_{{\bf k}\uparrow}^{L}$ and minority
$\epsilon_{{\bf k}\downarrow}^{L}$ subbands due to
the s-f (s-d) exchange.
The Hamiltonian $H_m^{L}$, Eq.~(\ref{hm}),
describes free magnons with spectrum $\omega_q$ which in the general case
has a gap $\omega_{q=0}=\omega_0$.
The third term $H_{em}^{L}$, Eq.~(\ref{hem}),
is the electron-magnon coupling resulting from the intra-atomic exchange
interaction between the spins of the conduction electrons
and the localized moments.

The calculation is performed using standard second order perturbation
theory.\cite{f+h65}
We write the total Hamiltonian, Eq.~(\ref{Htot}), as $H = H_0 + V$, 
where the perturbation $V = H_{T} + H_{em}^L$ is the sum
of the tunneling Hamiltonian and the electron-magnon interactions in the
left electrode.
First order terms provide an elastic contribution
to the current that do not involve any change of the spin orientation of the
itinerant electrons, whilst second order terms account for inelastic,
magnon-assisted processes.

%
%%%%%%%%%%%%%%%%%%%%%%%%%%%%%%%%%%%%%%%%%%%%%%%%%%%%%%%%%%%%%%%%%%%%
%%%%%%%%%%%%%%%%%%%%%%%%%%%%%%%%%%%%%%%%%%%%%%%%%%%%%%%%%%%%%%%%%%%%
\section{Current across a ferromagnetic-normal junction}\label{SECTcurrent}
%%%%%%%%%%%%%%%%%%%%%%%%%%%%%%%%%%%%%%%%%%%%%%%%%%%%%%%%%%%%%%%%%%%%
%%%%%%%%%%%%%%%%%%%%%%%%%%%%%%%%%%%%%%%%%%%%%%%%%%%%%%%%%%%%%%%%%%%%
%
%
%%%%%%%%%%%%%%%%%%%%%%%%%%%%%%%%%%%%%%%%%%%%%%%%%%%%%%%%%%%%%%%%%%%%
%%%%%%%%%%%%%%%%%%%%%%%%%%%%%%%%%%%%%%%%%%%%%%%%%%%%%%%%%%%%%%%%%%%%
\subsection{Elastic contribution to the current}
%%%%%%%%%%%%%%%%%%%%%%%%%%%%%%%%%%%%%%%%%%%%%%%%%%%%%%%%%%%%%%%%%%%%
%%%%%%%%%%%%%%%%%%%%%%%%%%%%%%%%%%%%%%%%%%%%%%%%%%%%%%%%%%%%%%%%%%%%
%
The first order contribution to the current arises from elastic tunneling
without any spin flip between either a spin up majority conduction electron
state on the left and a spin up state on the right or a spin down minority
state on the left and a spin down state on the right.
Consider for example an initial state consisting of an additional
majority spin up electron on the left with wavevector ${\bf k_L}$
and energy $\epsilon_{{\bf k_L}\uparrow}^{L}$.
This electron can tunnel, with matrix element $t_{LR}^{*}$,
into a spin up state on the right with
wavevector ${\bf k_R}$ and energy $\epsilon_{{\bf k_R}}^{R}$.
In addition there is a second process which is a transition between the same
two states, but in the reverse order, giving a contribution to the current
with an opposite sign.
Together, the two processes give a balance equation for the first order
contribution to the current between the spin up majority band on the left
and the spin up band on the right.
In addition, there are two first order processes that result in transitions
between the spin down minority band on the left and the spin down band on the
right.
Overall, the first order contribution to the current is $I_{\rm el}$ where
\begin{eqnarray}
I_{\rm el} &=& - 4 \pi^2 \frac{e}{h}
\int_{-\infty}^{+\infty} \!\! d\epsilon
\sum_{{\bf k_L}{\bf k_R}} \sum_{\alpha = \{ \uparrow , \downarrow \}}
\left| t_{{\bf k_L},{\bf k_R}} \right|^2
\delta (\epsilon - \epsilon_{{\bf k_L}\alpha}^L ) 
\nonumber \\
&& \times
\, \delta (\epsilon - eV - \epsilon_{{\bf k_R}}^R )
\Big\{
n_L (\epsilon_{{\bf k_L}\alpha}^L)
\left[ 1 - n_R (\epsilon_{{\bf k_R}}^R) \right] -
\nonumber \\
&& \qquad \qquad
- \left[ 1 - n_L (\epsilon_{{\bf k_L}\alpha}^L) \right]
n_R (\epsilon_{{\bf k_R}}^R)
\Big\} .
\label{i1a}
\end{eqnarray}
Keeping only contributions linear in $V$ or $\Delta T$,
the elastic contribution to the current may be written as
\begin{eqnarray}
I_{\rm el} &\approx& \frac{e^2}{h} V
\left[ {\cal T}_{\uparrow N} (\epsilon_F )
+ {\cal T}_{\downarrow N}  (\epsilon_F ) \right] \nonumber \\
&&- \left. \frac{\pi^2}{3} \frac{e}{h} (k_B^2 T \Delta T)
\frac{d}{d\epsilon} \left[ {\cal T}_{\uparrow N} (\epsilon )
+ {\cal T}_{\downarrow N}  (\epsilon ) \right] \right|_{\epsilon_F} .
\label{i1}
\end{eqnarray}
For convenience we have grouped all the information about the quality of the
interface into a parameter ${\cal T}_{\alpha N}$,
\begin{eqnarray}
{\cal T}_{\alpha N}  (\epsilon )
\approx
4 \pi^2 \sum_{{\bf k_L}{\bf k_R}}
\left| t_{{\bf k_L},{\bf k_R}} \right|^2
\delta (\epsilon - \epsilon_{\bf k_{L}\alpha}^L )
\delta (\epsilon - \epsilon_{\bf k_{R}}^R ) ,
\label{Tdef}
\end{eqnarray}
that is equivalent to the sum over all scattering channels,
from states with spin $\alpha$ on the left to states on the right
(where both spin channels are equivalent),
of the transmission eigenvalues usually introduced
in the Landauer formula,\cite{lan70,f+l81,bee97}
although we restrict ourselves to the tunneling regime in this paper.
Later we will employ models of two types of interface explicitly:
a uniformly transparent interface where the component of momentum
parallel to the interface
is conserved, and a randomly transparent interface.

The first term in Eq.~(\ref{i1}) accounts for the usual (large) contribution to
the electrical conductance,
\begin{eqnarray}
G_{\rm el} =  \frac{e^2}{h} [ {\cal T}_{\uparrow N} (\epsilon_F )
+ {\cal T}_{\downarrow N}  (\epsilon_F ) ] ,
\label{Gel}
\end{eqnarray}
and we define the corresponding degree of spin polarized current $P$ as the
relative difference in the current due to majority and minority carriers,
\begin{eqnarray}
P = \frac{\left( {\cal T}_{\uparrow N} - {\cal T}_{\downarrow N} \right)}
{\left( {\cal T}_{\uparrow N} + {\cal T}_{\downarrow N} \right)} .
\label{Pdef}
\end{eqnarray}
The thermopower coefficient $S_{\rm el} = -V/\Delta T$ due to the
elastic processes
may be determined from Eq.~(\ref{i1}) by setting $I_{\rm el} = 0$,
\begin{eqnarray}
S_{\rm el} \approx 
\left. - \frac{\pi^2}{3} \frac{k_B^2T}{e} 
\frac{d}{d\epsilon} \Big\{ \ln \left[ {\cal T}_{\uparrow N} (\epsilon )
+ {\cal T}_{\downarrow N}  (\epsilon ) \right] \Big\} \right|_{\epsilon_F} ,
\label{Sel}
\end{eqnarray}
which is equivalent to the Mott formula.\cite{mott}

%
%%%%%%%%%%%%%%%%%%%%%%%%%%%%%%%%%%%%%%%%%%%%%%%%%%%%%%%%%%%%%%%%%%%%
%%%%%%%%%%%%%%%%%%%%%%%%%%%%%%%%%%%%%%%%%%%%%%%%%%%%%%%%%%%%%%%%%%%%
\subsection{Magnon-assisted contribution to the current}
%%%%%%%%%%%%%%%%%%%%%%%%%%%%%%%%%%%%%%%%%%%%%%%%%%%%%%%%%%%%%%%%%%%%
%%%%%%%%%%%%%%%%%%%%%%%%%%%%%%%%%%%%%%%%%%%%%%%%%%%%%%%%%%%%%%%%%%%%

%
Below we describe processes which contribute to magnon-assisted tunneling.
We consider four processes, that are lowest order in the electron-magnon
interaction, as shown schematically in Figure~2.
The straight lines show the direction of electron transfer,
whereas the wavy lines denote the emission or absorption of magnons.
The processes are drawn using the rule, appropriate for ferromagnetic
electron-magnon exchange, that an electron in a minority state
scatters into a majority state by emitting a magnon.
The upper two processes in Figure~2, (i) and (ii), involve transitions
into (from) a majority final (initial) state on the left via
an intermediate, virtual state in the minority band.
For example, process (i), which is the same as the process
shown in more detail in Figure~1, begins with a
spin-down electron on the right with wavevector ${\bf k_R}$
and energy $\epsilon_{{\bf k_R}}^R$.
Then, this electron tunnels across the barrier (without spin flip) to occupy
a virtual, intermediate state with wavevector ${\bf k_L}$ in the spin down
minority band on the left as depicted in the left part of Figure~1(a)
with energy $\epsilon_{{\bf k_L}\downarrow}^L$.
The energy difference between the states is
$\epsilon_{{\bf k_L}\downarrow}^L - \epsilon_{{\bf k_R}}^R
= \epsilon_{{\bf k_L}\uparrow}^L - \epsilon_{{\bf k_R}}^R + \Delta$
so that the matrix element for the transition contains an energy in the
denominator related to the inverse lifetime of the electron in the virtual
state.
For $k_{B}T,eV \ll \Delta$, when both initial and final electron states should
be taken close to the Fermi level, only long wavelength magnons can be emitted,
so that the energy deficit in the virtual states can be approximated as
$\epsilon_{{\bf k_L}\uparrow}^L - \epsilon_{{\bf k_R}}^R + \Delta
\approx \Delta$.
As noticed in Refs.~\onlinecite{w+w70,mcc01},
this cancels out the large exchange parameter since the same electron-core
spin exchange constant appears both in the splitting between minority
and majority bands and in the electron-magnon coupling.
The second part of the transition is sketched in Figure~1(b) where
the electron in the virtual minority spin down state incorporates itself into
a state in the majority spin up band on the left,
wavevector ${\bf k^{\prime}}$, energy $\epsilon_{{\bf k^{\prime}}\uparrow}^L$,
by emitting a magnon of wavevector ${\bf q}$.

Similiar considerations enable us to write down the contribution to the current
from all the processes in Figure~2.
We group the processes into pairs which involve transitions between
the same series of states, but
in the opposite time order so that they give a current with different signs,
hence their sum gives a balance equation.
The contribution to the current of the processes 
(i) and (ii), labelled as $I_{\downarrow}$ because the state on
the right is spin down, is given by
\begin{eqnarray}
I_{\downarrow} &=& - 4 \pi^2 \frac{e}{h}
\int_{-\infty}^{+\infty} \!\! d\epsilon
\sum_{{\bf k^{\prime}}{\bf k_R}{\bf q}}
\frac{\left| t_{{\bf k^{\prime}},{\bf k_R}} \right|^2}{2\xi{\cal N}}
\nonumber \\
&& \times \,\delta (\epsilon - eV - \epsilon_{{\bf k_R}}^R )
\,\delta (\epsilon - \epsilon_{{\bf k^{\prime}}\uparrow}^L - \omega_{\bf q})
\nonumber \\
&& \times \Big\{
- n_R (\epsilon_{{\bf k_R}}^R)
\left[ 1 - n_L (\epsilon_{{\bf k^{\prime}}\uparrow}^L) \right]
\left[ 1 + N_L({\bf q}) \right] +
\nonumber \\
&& +
\left[ 1 - n_R (\epsilon_{{\bf k_R}}^R) \right]
n_L (\epsilon_{{\bf k^{\prime}}\uparrow}^L) N_L ({\bf q})
\Big\},
\label{iup}
\end{eqnarray}
where ${\bf q} = {\bf k_L} - {\bf k^{\prime}}$.
Energies on the right are shifted by $eV$ to take account of the
voltage difference across the junction.

%
%%%%%%%%%%%%%%%%%%%%%%%%%%%%%%%%%%%%%%%%%%%%%%%%%%%%%%%%%%%%%%%%%%%%%%%%%%
%%%%%%%%%%%%%%%%%%%%%%%%%%%%%%%%%%%%%%%%%%%%%%%%%%%%%%%%%%%%%%%%%%%%%%%%%%
%%%%%%%%%%%%%%%%%%%%%   figure 2 is inserted HERE %%%%%%%%%%%%%%%%%%%%%%%%
%%%%%%%%%%%%%%%%%%%%%%%%%%%%%%%%%%%%%%%%%%%%%%%%%%%%%%%%%%%%%%%%%%%%%%%%%%
%%%%%%%%%%%%%%%%%%%%%%%%%%%%%%%%%%%%%%%%%%%%%%%%%%%%%%%%%%%%%%%%%%%%%%%%%%
\begin{figure}
\epsfxsize=\hsize
\epsffile{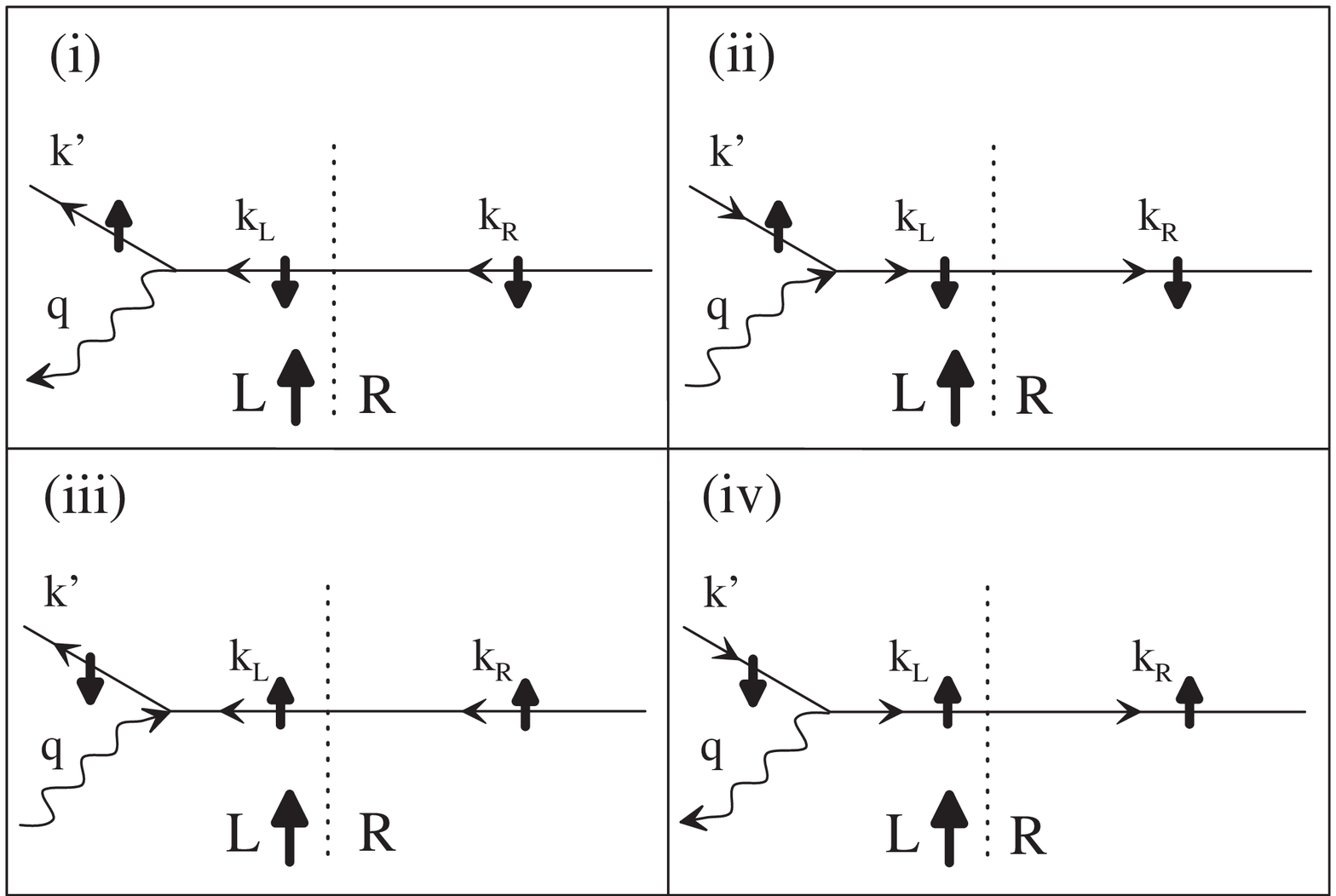}
\refstepcounter{figure}
\label{figure:2}

{\setlength{\baselineskip}{10pt} FIG.\ 2.
Schematic of the four magnon-assisted processes considered in the calculation.
The upper two processes, (i) and (ii), involve transitions
into (from) a majority final (initial) state on the left via
an intermediate, virtual state in the minority band
accompanied by magnon emission (absorption)
whereas the lower two processes, (iii) and (iv), involve transitions
into (from) a minority final (initial) state on the left via
an intermediate, virtual state in the majority band with
magnon absorption (emission).
}
\end{figure}
%%%%%%%%%%%%%%%%%%%%%%%%%%%%%%%%%%%%%%%%%%%%%%%%%%%%%%%%%%%%%%%%%%%%%%%%%%
%%%%%%%%%%%%%%%%%%%%%%%%%%%%%%%%%%%%%%%%%%%%%%%%%%%%%%%%%%%%%%%%%%%%%%%%%%
%%%%%%%%%%%%%%%%%%%%%   figure 2 is inserted above %%%%%%%%%%%%%%%%%%%%%%%
%%%%%%%%%%%%%%%%%%%%%%%%%%%%%%%%%%%%%%%%%%%%%%%%%%%%%%%%%%%%%%%%%%%%%%%%%%
%%%%%%%%%%%%%%%%%%%%%%%%%%%%%%%%%%%%%%%%%%%%%%%%%%%%%%%%%%%%%%%%%%%%%%%%%%
%

The lower two processes in Figure~2, (iii) and (iv), involve transitions
into (from) a minority final (initial) state on the left via
an intermediate, virtual state in the majority band from (into) a spin
up state on the right.
The contribution to the current from them, $I_{\uparrow}$,
is
\begin{eqnarray}
I_{\uparrow} &=& - 4 \pi^2 \frac{e}{h}
\int_{-\infty}^{+\infty} \!\! d\epsilon
\sum_{{\bf k^{\prime}}{\bf k_R}{\bf q}}
\frac{\left| t_{{\bf k^{\prime}},{\bf k_R}} \right|^2}{2\xi{\cal N}}
\nonumber \\
&& \!\!\! \times \,\delta (\epsilon - eV - \epsilon_{{\bf k_R}}^R )
\,\delta (\epsilon - \epsilon_{{\bf k^{\prime}}\downarrow}^L + \omega_{\bf q})
\nonumber \\
&& \!\!\! \times \Big\{
- n_R (\epsilon_{{\bf k_R}}^R)
\left[ 1 - n_L (\epsilon_{{\bf k^{\prime}}\downarrow}^L) \right]
N_L ({\bf q}) +
\nonumber \\
&& \!\! +
\left[ 1 - n_R (\epsilon_{{\bf k_R}}^R) \right]
n_L (\epsilon_{{\bf k^{\prime}}\downarrow}^L)
\left[ 1 + N_L({\bf q}) \right]
\Big\},
\label{idown}
\end{eqnarray}

We use the definition of the tunneling parameter ${\cal T}_{\alpha N}$,
Eq.~(\ref{Tdef}), in order to express the currents as 
\begin{eqnarray}
I_{\downarrow } &=& - \frac{e}{h}
\frac{{\cal T}_{\uparrow N}}{2\xi{\cal N}}
\int_{-\infty}^{+\infty} \!\! d\epsilon
\int_{0}^{\infty} \!\! d\omega \, \Omega (\omega ) \nonumber \\ 
&& \times \Big\{
N_L(\omega ) n_L(\epsilon ) \left[ 1 - n_R (\epsilon + \omega - eV ) \right]
\nonumber \\
&& \quad
- \left[ 1 + N_L (\omega ) \right]
\left[ 1 - n_L (\epsilon ) \right] n_R(\epsilon + \omega - eV) 
\Big\} ,
\label{iup2}
\end{eqnarray}
\begin{eqnarray}
I_{\uparrow } &=& + \frac{e}{h}
\frac{{\cal T}_{\downarrow N}}{2\xi{\cal N}}
\int_{-\infty}^{+\infty} \!\! d\epsilon
\int_{0}^{\infty} \!\! d\omega \, \Omega (\omega ) \nonumber \\ 
&& \times \Big\{
N_L(\omega ) n_R(\epsilon - \omega - eV ) \left[ 1 - n_L (\epsilon ) \right]
\nonumber \\
&& \quad
- \left[ 1 + N_L (\omega ) \right]
\left[ 1 - n_R(\epsilon - \omega - eV) \right] n_L (\epsilon )
\Big\} ,
\label{idown2}
\end{eqnarray}
where $\Omega (\omega ) = \sum_{\bf q} \delta (\omega - \omega_{{\bf q}})$
is the magnon density of states in the ferromagnet.
Since our main aim is to demonstrate the existence of an effect, we choose
the simple example of a bulk, three-dimensional magnon density of states.
We assume a quadratic magnon dispersion, $\omega_q = Dq^2$,
and apply the Debye approximation with a maximum magnon
energy $\omega_D = D (6 \pi^2 /v)^{2/3}$
where $v$ is the volume of a unit cell.
This enables us to express the magnon density of states as
$\Omega (\omega ) = (3/2) {\cal N} \omega^{1/2} / \omega_D^{3/2}$.

Since we are interested in the linear thermopower, we expand the electron
and magnon occupation numbers and keep only terms linear in $V$ and $\Delta T$. 
The resulting expression for the currents $I_{\downarrow }$ and
$I_{\uparrow }$,
Eqs.~(\ref{iup2}) and (\ref{idown2}) respectively, is 
\begin{eqnarray}
I_{\downarrow (\uparrow )} &=& \frac{e}{h}
\frac{3}{4\xi} \left( \frac{k_B T}{\omega_D} \right)^{3/2}
{\cal T}_{\uparrow (\downarrow ) N}  \times
\nonumber \\
&& \times
\Big\{
eV \, \Gamma (\textstyle\frac{5}{2})
\zeta (\textstyle\frac{3}{2})
\mp \textstyle\frac{1}{2} k_B\Delta T
\Gamma (\textstyle\frac{7}{2}) \zeta (\textstyle\frac{5}{2})
\Big\} .
\label{Iinel}
\end{eqnarray}
where $\Gamma (x)$ is the gamma function and
$\zeta (x)$ is Riemann's zeta function.\cite{g+r}

%
%%%%%%%%%%%%%%%%%%%%%%%%%%%%%%%%%%%%%%%%%%%%%%%%%%%%%%%%%%%%%%%%%%%%
%%%%%%%%%%%%%%%%%%%%%%%%%%%%%%%%%%%%%%%%%%%%%%%%%%%%%%%%%%%%%%%%%%%%
\section{Calculation of the thermopower}\label{SECTthermo}
%%%%%%%%%%%%%%%%%%%%%%%%%%%%%%%%%%%%%%%%%%%%%%%%%%%%%%%%%%%%%%%%%%%%
%%%%%%%%%%%%%%%%%%%%%%%%%%%%%%%%%%%%%%%%%%%%%%%%%%%%%%%%%%%%%%%%%%%%
%
The thermopower $S$ is determined by setting the total current to zero,
$I = I_{\rm el} + I_{\downarrow } + I_{\uparrow } = 0$,
and finding the voltage $V$ induced by the temperature
difference $\Delta T$, $S = - V/\Delta T$.
In the regime of temperatures given in Eq.~(\ref{reg0})
the total current may be approximated by
\begin{eqnarray}
I &\approx& \frac{e^2}{h} V
\left[ {\cal T}_{\uparrow N} + {\cal T}_{\downarrow N} \right] \nonumber \\
&& \!\!\! - \frac{e}{h}
\frac{3}{8\xi} \Gamma (\textstyle\frac{7}{2}) \zeta (\textstyle\frac{5}{2})
\left( \frac{k_B T}{\omega_D} \right)^{3/2}
k_B\Delta T
\left[ {\cal T}_{\uparrow N} - {\cal T}_{\downarrow N} \right] ,
\label{itot}
\end{eqnarray}
where the leading term proportional to $V$ arises from the elastic processes,
Eq.~(\ref{i1}), and the leading term proportional to $\Delta T$ comes from
the magnon-assisted processes, Eq.~(\ref{Iinel}).
The corresponding thermopower is
\begin{eqnarray}
S = - \frac{k_B}{e}
\frac{\delta m(T)}{2.077}
P \left( \Pi_{\uparrow (\downarrow )},\Pi_{N} \right) .
\label{Sall}
\end{eqnarray}
This is the main result of the paper, describing junctions between
normal metals and ferromagnets of arbitrary polarization strength ranging from
weak ferromagnets $\Pi_{\uparrow} \agt \Pi_{\downarrow}$
to half-metals $\Pi_{\uparrow} \gg \Pi_{\downarrow} = 0$.
The function $\delta m(T)$ in Eq.~(\ref{Sall}) is the change in the
magnetization due to thermal magnons at temperature $T$
(Bloch's $T^{3/2}$ law),\cite{kittel}
\begin{eqnarray}
\delta m(T)
&=& \frac{1}{\xi\cal{N}} \int_{0}^{\infty} d\omega \,
\Omega (\omega ) N_L (\omega )
\nonumber \\
&=&
\Gamma (\textstyle\frac{5}{2}) \zeta (\textstyle\frac{3}{2})\:\!
\frac{1}{\xi} \!\!\:
\left( \frac{k_BT}{\omega_D} \right)^{3/2} .
\label{mt}
\end{eqnarray}
The function $P$ appearing in the thermopower, Eq.~(\ref{Sall}), is the degree
of spin polarized current $P$ that flows in response to a bias voltage as
defined in Eq.~(\ref{Pdef}).
We consider two specific models of the interface: a uniformly transparent
interface where the component of momentum parallel to the interface
is conserved, and a randomly transparent interface.
As the form of the tunneling parameter ${\cal T}_{\alpha N}$,
Eq.~(\ref{Tdef}), has been given previously,\cite{m+f02b} we
only present the results here.
For a uniformly transparent interface of area $A$, such that the parallel
component of momentum is conserved upon tunneling, then
\begin{eqnarray}
{\cal T}_{\alpha N}^{\rm flat}
\approx 4 \pi^2 \left| t \right|^2
\frac{A}{h^2} \, {\rm min} \{ \Pi_{\alpha} , \Pi_{N} \} ,
\end{eqnarray}
where $t$ represents the transparency of the interface,
$\Pi_{\alpha}$ is the area of the maximal cross-section
of the Fermi surface of spins $\alpha$ in the ferromagnet,
$\Pi_{\uparrow} \geq \Pi_{\downarrow} \geq 0$,
and $\Pi_{N}$ is the area of the maximal cross-section
of the Fermi surface in the normal metal.
For a uniformly transparent interface,
\begin{eqnarray}
P^{\rm flat} &=&
\frac{\left( {\rm min} \{ \Pi_{\uparrow} , \Pi_{N} \} -
{\rm min} \{ \Pi_{\downarrow} , \Pi_{N} \} \right)}
{\left( {\rm min} \{ \Pi_{\uparrow} , \Pi_{N} \} +
{\rm min} \{ \Pi_{\downarrow} , \Pi_{N} \} \right)} .
\label{Pflat}
\end{eqnarray}
As an opposite extreme, we also consider a strongly nonuniform
interface which is transparent in a finite number of points only,
each of a typical area $a \sim \lambda_F^2$
randomly distributed over the interface area A.
In this case
\begin{eqnarray}
{\cal T}_{\alpha N}^{\rm dis}
&\approx& 4 \pi^2
\left| t \right|^2
\left( \frac{a\Pi_{\alpha}}{h^2} \right)
\left( \frac{a\Pi_{N}}{h^2} \right) , \\
P^{\rm dis} &=&
\frac{\left( \Pi_{\uparrow} - \Pi_{\downarrow} \right)}
{\left( \Pi_{\uparrow} + \Pi_{\downarrow} \right)}  .
\end{eqnarray}

The large thermopower Eq.~(\ref{Sall}) indicates that the current response
to $\Delta T$ of magnon-assisted processes is very efficient,
as compared to elastic processes.
We view this as being due to an attempt by the bath of magnons to alter its
temperature.
Since the population of magnons defines the magnetization of the ferromagnet,
through the function $\delta m(T)$ Eq.~(\ref{mt}),
thermal equilibration achieved by changing the magnon population must
be accompanied by a change in magnetization.
This is mediated by conduction electrons, resulting in a net current
response to a temperature difference across the junction
Eqs.~(\ref{Iinel}) and~(\ref{itot}).
For example, the process shown in Figure~2(ii), in which an electron on the
left absorbs a magnon and tunnels to the right, results in a reduction in
the number of magnons and the injection of a spin down electron to the right.
On the other hand, the process shown in Figure~2(iii), in which an electron on
the right tunnels to the left and absorbs a magnon, results in a reduction in
the number of magnons and the collection of a spin up electron from the right.
These two processes, while both lowering the number of magnons, give competing
contributions to the current as demonstrated by the factor
${\cal T}_{\uparrow N} - {\cal T}_{\downarrow N}$ in Eq.~(\ref{itot}).

%
%%%%%%%%%%%%%%%%%%%%%%%%%%%%%%%%%%%%%%%%%%%%%%%%%%%%%%%%%%%%%%%%%%%%
%%%%%%%%%%%%%%%%%%%%%%%%%%%%%%%%%%%%%%%%%%%%%%%%%%%%%%%%%%%%%%%%%%%%
\section{Conclusion}\label{SECTconc}
%%%%%%%%%%%%%%%%%%%%%%%%%%%%%%%%%%%%%%%%%%%%%%%%%%%%%%%%%%%%%%%%%%%%
%%%%%%%%%%%%%%%%%%%%%%%%%%%%%%%%%%%%%%%%%%%%%%%%%%%%%%%%%%%%%%%%%%%%
%
The main result of this work is a large contribution to the thermopower
Eq.~(\ref{Sall}) due to magnon-assisted processes in ferromagnetic-normal
metal tunnel junctions.
As a rough estimate, we take
$\delta m = 7.5 \!\times \! 10^{-6} \, T^{3/2}$
(for a ferromagnet such as Ni, Ref.~\onlinecite{kittel})
and $P \sim 0.4$ to give $S \sim - 1 {\rm\mu \;\!\! V\,K^{-1}}$ at $T = 300$K.
Our simple model describes tunneling between parabolic conduction bands
typical of three dimensional metallic systems, with additional spin
splitting and electron-magnon interactions in the ferromagnet.
However we believe the main results will have
qualitative relevance for junctions with semiconducting elements, too.
Recent numerical modeling of the diluted magnetic semiconductor
(Ga,Mn)As using a six-band Kohn-Luttinger Hamiltonian\cite{kon01}
found evidence of quadratic dispersion of long-wavelength spin waves
$\omega_q = \omega_0 + Dq^2$ with a small anisotropy gap $\omega_0$.
A fit at small momenta to their data for typical sample parameters
yields a spin waves stiffness $D \sim 2$ meV nm$^2$ that corresponds
to $\delta m \sim 2.5 \!\times \! 10^{-4} \, T^{3/2}$.
If this is inserted into our formula for the thermopower Eq.~(\ref{Sall}) with
$P \sim 0.4$, say, it gives
$|S| \sim 4 {\rm\mu \;\!\! V\,K^{-1}}$ at $T = 100$K.
However, we stress that the sign of the thermopower Eq.~(\ref{Sall}) is
specified for electron (charge $-e$) transfer processes between parabolic
conduction bands and under the assumption that the exchange between
conduction band and core electrons has a ferromagnetic sign.
We considered a bulk, three-dimensional magnon density of states,
but in general the magnitude and sign of the thermopower will depend
on details of the magnon spectrum.

%%%%%%%%%%%%%%%%%%%%%%%%%%%%%%%%%%%%%%%%%%%%%%%%%%%%%%%%%%%%%%%%%%%%
%%%%%%%%%%%%%%%%%%%%%%%%%%%%%%%%%%%%%%%%%%%%%%%%%%%%%%%%%%%%%%%%%%%%
 
The authors thank J.~F.~Annett, B.~L.~Gyorffy, T.~Jungwirth,
A.~H.~MacDonald, and G.~Tkachov for discussions,
and EPSRC for financial support.

%%%%%%%%%%%%%%%%%%%%%%%%%%%%%%%%%%%%%%%%%%%%%%%%%%%%%%%%%%%%%%%%%%%%
%%%%%%%%%%%%%%%%%%%%%%%%%%%%%%%%%%%%%%%%%%%%%%%%%%%%%%%%%%%%%%%%%%%%

\end{multicols}
%%%%%%%%%%%%%%%%%%%%%%%%%%%%%%%%%%%%%%%%%%%%%%%%%%%%%%%%%%%%%%%%%%%%%

\end{document}